\documentclass[12pt]{article}

\ifx\pdfoutput\undefined
\usepackage[dvips,bookmarks]{hyperref}
\else
\usepackage{hyperref}
\fi
\hypersetup{colorlinks=false,bookmarksopen,bookmarksnumbered,citecolor=blue,
   pdfstartview=FitH}

\usepackage[dvips]{graphicx}
\usepackage{latexsym}

\usepackage{amssymb,amsfonts,amsmath}
\usepackage{graphicx} 
\usepackage{indentfirst}

 \usepackage{bbm}

\parskip=\medskipamount

\newcommand {\cD}{{\cal D}}
\newcommand {\cE}{{\cal E}}

\newcommand {\cK}{{\cal K}}
\newcommand {\cL}{{\cal L}}
\newcommand {\cM}{{\cal M}}
\newcommand {\cN}{{\cal N}}

\newcommand {\cS}{{\cal S}}

\newcommand {\cU}{{\cal U}}

\def\a{\alpha}
\def \bi{\bibitem}

\def\b{\beta}

\def\d{\delta}

\def\g{\gamma}

\def\k{\kappa}
\def\l{\lambda}
\def\m{\mu}

\def\q{\theta}

\def\s{\sigma}

\def\u{\upsilon}
\def\x{\xi}
\def\z{\zeta}

\def\F{\Phi}

\def\L{\Lambda}
\def\O{\Omega}

\def\Q{\Theta}
\def\S{\Sigma}
\def\U{\Upsilon}

\def\rd{{\rm d}}
\def\ri{{\rm i}}

\newcommand{\ad}{{\dot{\alpha}}}                           
\newcommand{\bd}{{\dot{\beta}}}                            
\newcommand{\ve}{\varepsilon}                            
\newcommand{\cDB}{{\bar\cD}}                            

\newcommand{\pa}{\partial}                           
\newcommand{\hf}{\frac12}

\newcommand{\be}{\begin{equation}}
\newcommand{\ee}{\end{equation}}
\newcommand{\bea}{\begin{eqnarray}}
\newcommand{\eea}{\end{eqnarray}}
\newcommand{\non}{\nonumber}
\newcommand{\1}{\underline{1}}
\newcommand{\2}{\underline{2}}

\newcommand{\bm}[1]{\mbox{\boldmath$#1$}}

\def\double #1{#1{\hbox{\kern-2pt $#1$}}}

\newcommand{\gd}{{\dot\g}}
\newcommand{\dd}{{\dot\d}}

\newcommand{\ts}{{\tilde{\s}}}

\newcommand{\sba}{{\bar{\s}}}

\newcommand{\teb}{{\bar{\theta}}}

\begin{document}
\begin{titlepage}
\begin{flushright}
UUITP-07/08\\
YITP-SB-08-27\\
May, 2008\\
\end{flushright}
\vspace{5mm}

\begin{center}
{\Large \bf  4D  $\bm{\cN=2}$ Supergravity and Projective Superspace}\\ 
\end{center}

\begin{center}

{\bf
S. M. Kuzenko\footnote{kuzenko@cyllene.uwa.edu.au}${}^{a}$,
U. Lindstr\"om\footnote{ulf.lindstrom@teorfys.uu.se}${}^{b}$,
M. Ro\v{c}ek\footnote{rocek@max2.physics.sunysb.edu}${}^{c}$, 
G. Tartaglino-Mazzucchelli\footnote{gtm@cyllene.uwa.edu.au}${}^{a}$
} \\
\vspace{5mm}

\footnotesize{
${}^{a}${\it School of Physics M013, The University of Western Australia\\
35 Stirling Highway, Crawley W.A. 6009, Australia}}  
~\\
\vspace{2mm}

\footnotesize{
${}^{b}${\it Department of Theoretical Physics,
Uppsala University \\ 
Box 803, SE-751 08 Uppsala, Sweden}\\

 and\\
 
{\it  NORDITA, Roslagstullsbacken 23,\\
 SE-10691 Stockholm, Sweden
}}
\\
\vspace{2mm}

${}^c${\it C.N.Yang Institute for Theoretical Physics, Stony Brook University \\
Stony Brook, NY 11794-3840,USA}

\end{center}
\vspace{5mm}

\begin{abstract}
\baselineskip=14pt
This paper presents a projective superspace formulation for 4D $\cN=2$ 
matter-coupled supergravity. We first describe a variant superspace realization
for the $\cN=2$ Weyl multiplet. 
It differs from that proposed by Howe 
in 1982 by the choice of the structure group
\big(${\rm SO(3,1)} \times  {\rm SU(2)}$ versus 
${\rm SO(3,1)} \times  {\rm U(2)} $\big), which implies that the super-Weyl   
transformations are generated by a covariantly chiral parameter instead of a real
unconstrained one.
We introduce various off-shell supermultiplets  
which are curved superspace analogues of the superconformal projective 
multiplets in global supersymmetry 
and which describe matter fields coupled to supergravity.
A manifestly locally supersymmetric and super-Weyl invariant action principle is 
given. 
Off-shell locally supersymmetric nonlinear sigma models are presented in this new superspace.
\end{abstract}
\vspace{1cm}

\vfill
\end{titlepage}

\newpage
\renewcommand{\thefootnote}{\arabic{footnote}}
\setcounter{footnote}{0}


\section{Introduction}
\setcounter{equation}{0}

The increasing number of spinor derivatives in the superspace measure in theories with
higher supersymmetry is 
a well-known obstacle to the construction of extended superspace actions.
A resolution of the problem lies in finding invariant subspaces over which to integrate. 
One such setting is four-dimensional $\cN=2$ projective superspace 
\cite{KLR} (a related method uses the harmonic superspace\footnote{Both 
methods make use of the superspace 
${\mathbb R}^{4|8}\times {\mathbb C}P^1={\mathbb R}^{4|8}\times S^2$ 
introduced for the first time in \cite{Rosly}.
However, they differ in  (i) the structure of off-shell supermultiplets used; 
and (ii) the supersymmetric action principle chosen. 
Due to these conceptual differences, the two approaches prove to be  complementary 
to each other in many respects.
The relationship between the harmonic and projective superspace formulations
is spelled out in \cite{K-double}.}
of, {\it e.g.}, \cite{harm1,harm}). 
Its applications 
include classical sigma models and their quantization \cite{G-RLRvUW},  
as well as supersymmetric Yang-Mills theory \cite{Lindstrom:1989ne}, \cite{GonzalezRey:1997db}. 
In particular, the sigma model description is well suited for the construction of 
new hyperk\"ahler metrics \cite{LR}, \cite{Arai:2007cx}.  
The projective supermultiplets \cite{KLR,Lindstrom:1989ne,LR,KLT} 
and the action principle \cite{KLR} are at the heart of this approach.  
${}$For the quantum theory, it is imperative to have an off-shell formalism and 
extremely useful to have all symmetries manifest. Geometrically, projective superspace 
is closely connected to twistor space, a property which is being extensively studied \cite{infty}.

The concept of projective superspace has also proven to be very useful for 
supersymmetric theories with eight supercharges in 
five \cite{KL} and six  \cite{GL,GPT}  dimensions. 
Superconformal field theory in projective superspace 
has been developed in four and five dimensions \cite{K,K2}, 
including the formulation of general off-shell superconformal sigma models.

What has been lacking in the formalism is a description of supergravities 
with eight supercharges in diverse dimensions. 
Recently this problem has been solved 
in the case of five-dimensional $\cN=1$ matter-coupled supergravity
\cite{KT-M2,KT-M3,KT-M4}.
In the present paper we develop
a projective superspace formulation for
four-dimensional ${\cal N}=2$ supergravity 
and its matter couplings. 
In particular, we identify a suitable set of  constraints
which are compatible with a super-Weyl invariance
and  with  the existence of a large family of local projective multiplets, 
{\it i.e.}, curved space versions of the superconformal projective multiplets \cite{K2}. 
This allows us to elaborate a 
conformal supergravity setting for general 
$\cN=2$ supergravity-matter systems 
similar
to that existing in the $\cN=1$ case  
as reviewed in, {\it e.g.}, \cite{Gates:1983nr,BK}. 
Our results also include the coupling of 
the conformal supergravity to vector multiplets, a super Weyl-invariant 
action for supergravity-matter systems in projective superspace, 
and new formulations of off-shell locally supersymmetric nonlinear sigma models. 

The paper is organized as follows:  In section 2 we introduce (two) superspace formulations of the 
Weyl multiplet using Grimm's constraints and solution
for ${\cal N}=2$ supergravity \cite{Grimm}, and comment briefly on the relation to the superspace formulation of Howe \cite{Howe}. In section 3 we define the relevant projective multiplets and their transformations. Section 4 contains the coupling of the conformal supergravity multiplet to vector supermultiplets and in section 5 we formulate a locally supersymmetric and 
super-Weyl-invariant  action in which the Lagrangian is
a real projective multiplet of weight two coupled to conformal supergravity.

Before turning to the technical part of this paper, it is worth comparing 
the current status of superspace approaches to general matter couplings  
in $\cN=2$ supergravity to those developed long ago for $\cN=1$ supergravity. 
In the latter case, there exist two main formalisms: 
(i) the constrained  geometric formulation mainly due to Wess and Zumino \cite{WZ-s};
(ii) the unconstrained prepotential formulation which was presented in the most elaborated form 
in \cite{SG}. The approaches (i) and (ii) are intimately related, since the prepotential formulation 
is obtained by solving the Wess-Zumino constraints in terms of unconstrained 
superfields. In the case of $\cN=2$ supergravity, a prepotential formulation
was proposed within the harmonic superspace approach twenty years ago 
\cite{Galperin:1987em,Galperin:1987ek}. However, 
a  relationship of this prepotential formulation to 
the standard curved $\cN=2$  superspace geometry has never been elaborated in detail
(except one specific off-shell realization for $\cN=2$ Poincar\'e supergravity
considered in \cite{Galperin:1987em}).
In particular, it has never been shown how the harmonic prepotentials introduced in 
\cite{Galperin:1987ek} occur as a result of  
solving the constraints in Howe's formulation for $\cN=2$ conformal supergravity \cite{Howe}.
On the other hand, the recent study  of 5D $\cN=1$ supergravity
\cite{KT-M2,KT-M3,KT-M4} and the present paper
clearly demonstrate that 
projective superspace is ideal  for developing covariant geometric formulations 
for supergravity-matter systems  with eight supercharges.
Ultimately, a completely coherent superspace description of $\cN=2$ supergravity 
should probably require a synthesis of the harmonic and projective superspace methods. 
Keeping this in mind, we  intentionally use in this paper 
a notation consistent with both approaches.

\section{Variant formulations for the ${\cN=2}$ Weyl multiplet}
\setcounter{equation}{0}

The Weyl multiplet of four-dimensional $\cN=2$ conformal supergravity 
 \cite{deWvHVP,BdeRdeW,deWLVP}  
 was realized in superspace long ago by Howe \cite{Howe}
 (see also \cite{Gates:1980ky} for the earlier discussion of the superconformal 
 aspects of $\cN=2$ supergravity in superspace).
The structure group in his approach is chosen to be 
${\rm SO(3,1)} \times  {\rm SU(2)} \times{\rm U(1)}$, 
and the super-Weyl transformations are generated by a real unconstrained parameter.
We have not found the formulation given in \cite{Howe} to be the simplest 
from the point of view of the explicit off-shell construction of  supergravity-matter systems.
Here we present an alternative superspace formulation for ${\cN=2}$ conformal supergravity.
It differs from that given in  \cite{Howe} in the following three points: 
(i) the structure group is identified with 
 ${\rm SO(3,1)} \times  {\rm SU(2)}$; (ii) the geometry of curved superspace is
 subject to the constraints introduced by Grimm \cite{Grimm}; 
 (iii)  the  super-Weyl transformations 
 are generated by a covariantly chiral but otherwise unconstrained superfield.
 We will briefly discuss the  precise correspondence between 
 the two formulations for conformal supergravity  at the end of this section.

\subsection{Grimm's superspace geometry}
Consider a curved 4D $\cN=2$ superspace  $\cM^{4|8}$ parametrized by 
local bosonic ($x$) and fermionic ($\q, \bar \q$) 
coordinates  $z^{{M}}=(x^{m},\q^{\mu}_i,{\bar \q}_{\dot{\mu}}^i)$,
where $m=0,1,\cdots,3$, $\mu=1,2$, $\dot{\mu}=1,2$ and  $i=\1,\2$.
The Grassmann variables $\q^{\mu}_i $ and $\teb_{\dot{\mu}}^i$
are related to each other by complex conjugation: 
$\overline{\q^{\mu}_i}=\teb^{\dot{\mu}i}$. 
Following \cite{Grimm},
the structure group is chosen to be ${\rm SO}(3,1)\times {\rm SU}(2)$,
and the covariant derivative 
$\cD_{{A}} =(\cD_{{a}}, \cD_{{\a}}^i,\cDB^\ad_i)$
have the form 
\bea
\cD_{A}&=&E_{A}~+~
\Phi_{A}^{~\,kl}\,J_{kl}~+~\hf \,\O_{A}{}^{bc}\,M_{bc}\non\\
&=&E_{A}~+~\Phi^{~\,kl}_{A}\,J_{kl}~+~\O_{A}{}^{\b\g}\,M_{\b\g}
+{\bar \O}_{A}{}^{\bd\gd}\,\bar{M}_{\bd\gd}
~.
\label{CovDev}
\eea
Here $E_{{A}}= E_{{A}}{}^{{M}}(z) \pa_{{M}}$ is the supervielbein, 
with $\pa_{{M}}= \pa/ \pa z^{{M}}$,
$J_{kl}=J_{lk}$
are generators of the group SU(2), 
$M_{bc}$ the generators of the Lorentz group ${\rm SO}(3,1)$, 
 $\Phi_{{A}}{}^{kl}(z)$  and $\O_{{A}}{}^{bc}(z)$ the corresponding connections. 
The Lorentz generators with vector indices ($M_{ab}=-M_{ba}$) and spinor indices
($M_{\a\b}=M_{\b\a}$ and ${\bar M}_{\ad\bd}={\bar M}_{\bd\ad}$) are related to each other 
by the rule:
$$
M_{ab}=(\s_{ab})^{\a\b}M_{\a\b}-(\tilde{\s}_{ab})^{\ad\bd}\bar{M}_{\ad\bd}~,~~~
M_{\a\b}=\hf(\s^{ab})_{\a\b}M_{ab}~,~~~
\bar{M}_{\ad\bd}=-\hf(\tilde{\s}^{ab})_{\ad\bd}M_{ab}~.
$$ 
The generators of the structure group
act on the covariant derivatives as follows:\footnote{In what follows, 
the (anti)symmetrization of $n$ indices 
is defined to include a factor of $(n!)^{-1}$.}
\bea
{[}J_{kl},\cD_{\a}^i{]}
&=&-\d^i_{(k}\cD_{\a l)}~,
\qquad
{[}J_{kl},\cDB^{\ad}_i{]}
=-\ve_{i(k}\cDB^\ad_{l)}~, \non \\
{[}M_{\a\b},\cD_{\g}^i{]}
&=&\ve_{\g(\a}\cD^i_{\b)}~,\qquad
{[}\bar{M}_{\ad\bd},\cDB_{\gd}^i{]}=\ve_{\gd(\ad}\cDB^i_{\bd)}~,
\label{generators}
\eea
while 
${[}M_{\a\b},\cDB_{\gd}^i{]}=
{[}\bar{M}_{\ad\bd},\cD_{\g}^i{]}=0.$
Our notation and conventions correspond to \cite{BK}; they 
almost coincide with 
those used in \cite{WB} except for the normalization of the 
Lorentz generators, including a sign definition of  
the sigma-matrices $\s_{ab}$ and $\tilde{\s}_{ab}$.

The supergravity gauge group is generated by local transformations
of the form 
\be
\d_K \cD_{{A}} = [K  , \cD_{A}]~,
\qquad K = K^{{C}}(z) \cD_{{C}} +\hf K^{ c  d}(z) M_{c  d}
+K^{kl}(z) J_{kl}  ~,
\label{tau}
\ee
with the gauge parameters
obeying natural reality conditions, but otherwise  arbitrary. 
Given a tensor superfield $U(z)$, with its indices suppressed, 
it transforms as follows:
\bea
\d_K U = K\, U~.
\label{tensor-K}
\eea

The  covariant derivatives obey (anti-)commutation relations of the form:
\bea
{[}\cD_{{A}},\cD_{{B}}\}&=&T_{{A}{B}}{}^{{C}}\cD_{{C}}
+R_{{A}{B}}{}^{kl}J_{kl}
+\hf R_{{A}{B}}{}^{{c}{d}}M_{{c}{d}}
~,
\label{algebra}
\eea
where $T_{{A}{B}}{}^{{C}}$ is the torsion, and $R_{{A}{B}}{}^{kl}$ and $R_{{A}{B}}{}^{{c}{d}}$ 
constitute the curvature.
${}$Following  \cite{Grimm},  some components of the torsion are subject to the 
following constraints:
\begin{subequations}
\bea
T_{\a}^i{}^\bd_j{}^c=-2\ri\d^i_j(\s^{c})_{\a}{}^{\bd}~,\qquad
T_{\a}^i{}_\b^j{}^{c}=T{}^\ad_i{}^\bd_j{}^{c}=0 \quad && \quad
\mbox{(dim 0)}
\label{constr-0}\\
T_\a^i{}_\b^j{}^\g_k=T_\a^i{}_\b^j{}_\gd^k
=T_\a^i{}^\bd_j{}^\g_k=T_\a^i{}^\bd_j{}_\gd^k
=T{}^\ad_i{}^\bd_j{}_\gd^k=
T_\a^i{}_b{}^c=T{}^\ad_i{}_b{}^c=0
\quad && \quad
\mbox{(dim $\frac{1}{2}$)}
\label{constr-1/2}\\
T_{a}{}_{\b}^j{}^\g_k=\d^j_k\,T_{a\b}{}^\g~,\qquad 
T_{a}{}^\bd_j{}_\gd^k=\d_j^k\,T_a{}^\bd{}_\gd~,\qquad
T_{ab}{}^{c}=0
\quad && \quad
\mbox{(dim 1)}~.
\label{constr-1}
\eea
\end{subequations}
The solution to the constraints was given in \cite{Grimm}. 
Modulo trivial re-definitions, it is: 
\begin{subequations} 
\bea
\{\cD_\a^i,\cD_\b^j\}&=&
4S^{ij}M_{\a\b}
+2\ve^{ij}\ve_{\a\b}Y^{\g\d}M_{\g\d}
+2\ve^{ij}\ve_{\a\b}\bar{W}^{\gd\dd}\bar{M}_{\gd\dd}
\non\\
&&
+2 \ve_{\a\b}\ve^{ij}S^{kl}J_{kl}
+4 Y_{\a\b}J^{ij}~, 
\label{acr1} \\
\{\cDB^\ad_i,\cDB^\bd_j\}&=&
-4\bar{S}_{ij}\bar{M}^{\ad\bd}
-2\ve_{ij}\ve^{\ad\bd}\bar{Y}^{\gd\dd}\bar{M}_{\gd\dd}
-2\ve_{ij}\ve^{\ad\bd}{W}^{\g\d}M_{\g\d}
\non\\
&&
-2\ve_{ij}\ve^{\ad\bd}\bar{S}^{kl}J_{kl}
-4\bar{Y}^{\ad\bd}J_{ij}~,
\label{acr2} \\
\{\cD_\a^i,\cDB^\bd_j\}&=&
-2\ri\d^i_j(\s^c)_\a{}^\bd\cD_c
+4\d^{i}_{j}G^{\d\bd}M_{\a\d}
+4\d^{i}_{j}G_{\a\gd}\bar{M}^{\gd\bd}
+8 G_\a{}^\bd J^{i}{}_{j}~,
\eea
\bea
{[}\cD_a,\cD_\b^j{]}&=&
\ri(\s_a)_{(\b}{}^{\bd}G_{\g)\bd}\cD^{\g j}
+{\frac{\ri}2}\Big(({\s}_a)_{\b\gd}S^{jk}
-\ve^{jk}({\s}_a)_\b{}^{\dd}\bar{W}_{\dd\gd}
-\ve^{jk}({\s}_a)^{\a}{}_\gd Y_{\a\b}\Big)\cDB^\gd_k
\non\\
&&
+{\frac{\ri}2}\Big((\s_a)_{\b}{}^{\dd} T_{cd}{}_\dd^j
+(\s_c)_{\b}{}^{\dd} T_{ad}{}_\dd^j
-(\s_d)_{\b}{}^{\dd} T_{ac}{}_\dd^j\Big)M^{{c}{d}}
\non\\
&&
+{\frac{\ri}2}\Big((\ts_a)^{\gd\g}\ve^{j(k}\cDB_\gd^{l)}Y_{\b\g}
-(\s_a)_{\b\gd}\ve^{j(k}\cDB_{\dd}^{l)}\bar{W}^{\gd\dd}
-{\frac12}(\s_a)_\b{}^{\gd}\cDB_{\gd}^{j}S^{kl}\Big)
J_{kl}~,\\
{[}\cD_a,\cDB^\bd_j{]}&=&
-\ri(\s_a)_\a{}^{(\bd}G^{\a\gd)}
\cDB_{\gd j}
+{\frac{\ri}2}
\Big(({\ts}_a)^{\bd\g}\bar{S}_{jk}
-\ve_{jk}({\s}_a)_\a{}^{\bd}{W}^{\a\g}
-\ve_{jk}({\s}_a)^{\g}{}_\ad\bar{Y}^{\ad\bd}\Big)
\cD_\g^k
\non\\
&&
+{\frac{\ri}2}\Big((\ts_a)_\d{}^{\bd}T_{cd}{}^{\d}_{ j}
+(\s_c)_\d{}^{\bd}T_{ad}{}^{\d}_{ j}
-(\s_d)_\d{}^{\bd}T_{ac}{}^{\d}_{ j}\Big)
M^{cd}
\non\\
&&
+{\frac{\ri}2}
\Big(-(\s_a)^{\g}{}_\gd\d_{j}^{(k}\cD_{\g}^{l)}\bar{Y}^{\bd\gd}
-(\s_a)_{\g}{}^{\bd}\d_{j}^{(k}\cD_{\d}^{ l)}{W}^{\g\d}
+{\frac12}(\s_a)_\a{}^\bd\cD^{\a}_{ j}\bar{S}^{kl}\Big)
J_{kl}~,~~~
\eea
\end{subequations}
where 
\begin{subequations}
\bea
T_{ab}{}^\g_k&=&-{\frac14}(\ts_{ab})^{\ad\bd}\cD^\g_k\bar{Y}_{\ad\bd}
+{\frac14}(\s_{ab})^{\a\b}\cD^\g_kW_{\a\b}
-{\frac16}(\s_{ab})^{\g\d}\cD^l_\d\bar{S}_{kl}~,\\
T_{ab}{}_\gd^k&=&-{\frac14}(\s_{ab})^{\a\b}\cDB_\gd^{ k}Y_{\a\b}
+{\frac14}(\ts_{ab})^{\ad\bd}\cDB_{\gd}^{ k}\bar{W}_{\ad\bd}
-{\frac16}(\ts_{ab})_{\gd\dd}\cDB^{\dd}_{l}S^{kl}~.
\eea
\end{subequations}
Here the real four-vector $G_{\a \ad} $,
the complex symmetric  tensors $S^{ij}=S^{ji}$, $W_{\a\b}=W_{\b\a}$, 
$Y_{\a\b}= Y_{\b\a}$ and their complex conjugates 
$\bar{S}_{ij}:=\overline{S^{ij}}$, $\bar{W}_{\ad\bd}:=\overline{W_{\a\b}}$,
$\bar{Y}_{\ad\bd}:=\overline{Y_{\a\b}}$ obey additional constraints implied 
by the Bianchi identities.
They comprise the  dimension 3/2 identities 
\begin{subequations}
\bea
\cD_{\a}^{(i}S^{jk)}= {\bar \cD}_{\ad}^{(i}S^{jk)}&=&0~,
\label{S-analit}\\
\cDB^\ad_iW_{\b\g}&=&0~,\qquad\\
\cD_{(\a}^{i}Y_{\b\g)}&=&0~,\\
\cD_{\a}^{i}S_{ij}+\cD^{\b}_{j}Y_{\b\a}&=&0~, \\
\cD_\a^iG_{\b\bd}&=&
-{\frac14}\cDB_{\bd}^iY_{\a\b}
+\frac{1}{12}\ve_{\a\b}\cDB_{\bd j}S^{ij}
-{\frac14}\ve_{\a\b}\cDB^{\gd i}\bar{W}_{\bd\gd}~,
\eea
\end{subequations}
as well as the dimension 2 relation
\bea
\big( \cD^i_{(\a} \cD_{\b) i}
-4Y_{\a\b} \big) W^{\a\b}
&=& \big( \cDB_i^{( \ad }\cDB^{\bd ) i}
-4\bar{Y}^{\ad\bd} \big) \bar{W}_{\ad\bd}
~.
\label{dim-2-constr}
\eea
At this point, Grimm stopped his analysis in 1980 \cite{Grimm}.

It is worth pointing out that the 4D $\cN=2$ anti-de Sitter superspace 
$$
\frac{{\rm OSp}(2|4)}{{\rm SO}(3,1) \times {\rm SO} (2)}
$$
corresponds to a supergeometry with covariantly constant torsion
(compare with the case of 5D $\cN=1$ anti-de Sitter superspace \cite{KT-M}):
\be
W_{\a\b}=Y_{\a\b}=0~, 
\qquad G_{\a \bd}=0~, \qquad 
\cD^i_\a S^{kl} = {\bar \cD}^i_\ad S^{kl}=0~.
\ee
The integrability condition for these constraints 
is $[S, S^\dagger ]=0$, with $S= (S^i{}_j)$,
and hence ${\bar S}^{ij} = q \, \cS^{ij}$, where 
$ {\bar \cS}^{ij}= \cS^{ij}$ and $q \in $ U(1) is a constant parameter.

\subsection{Super-Weyl transformations}

What was not noticed in \cite{Grimm}, is the fact 
that the constraints (\ref{constr-0} -- \ref{constr-1})
are invariant under super-Weyl transformations of the form:
\bea
\d_{\s} \cD_\a^i&=&\hf\sba\cD_\a^i+(\cD^{\g i}\s)M_{\g\a}-(\cD_{\a k}\s)J^{ki}~, \non\\
\d_{\s} \cDB_{\ad i}&=&\hf\s\cDB_{\ad i}+(\cDB^{\gd}_{i}\sba)\bar{M}_{\gd\ad}
+(\cDB_{\ad}^{k}\sba)J_{ki}~, 
\label{super-Weyl1} \\
\d_{\s} \cD_a&=&
\hf(\s+\sba)\cD_a
+{\frac{\ri}4}(\s_a)^\a{}_{\bd}(\cD_{\a}^{ k}\s)\cDB^{\bd}_{ k}
+{\frac{\ri}4}(\s_a)^{\a}{}_\bd(\cDB^{\bd}_{ k}\sba)\cD_{\a}^{ k}
-{\frac12}\big(\cD^b(\s+\sba)\big)M_{ab}
~, \non
\eea
where  the parameter $\s$ is an arbitrary covariantly chiral superfield, 
\be
{\bar \cD}_{\ad i} \s=0~.
\ee
The  dimension-1 components of the torsion transform under 
(\ref{super-Weyl1}) as follows:
\begin{subequations}
\bea
\d_{\s} S^{ij}&=&\sba S^{ij}-{\frac14}\cD^{\g(i}\cD^{j)}_\g \s~, 
\label{super-Weyl-S} \\
\d_{\s} Y_{\a\b}&=&\sba Y_{\a\b}-{\frac14}\cD^{k}_{(\a}\cD_{\b)k}\s~,
\label{super-Weyl-Y} \\
\d_{\s} {W}_{\a \b}&=&\s {W}_{\a \b }~,\\
\d_{\s} G_{\a\bd} &=&
\hf(\s+\sba)G_{\a\bd} -{\frac{\ri}4}
\cD_{\a \bd} (\s-\sba)~.
 \label{super-Weyl-G} 
\eea
\end{subequations}
Observe that the covariantly chiral bi-spinor $W_{\a\b}$ transforms homogeneously, 
and therefore it is a superfield extension of the Weyl tensor, and 
that the $\q$-independent component of $G_a$, 
$V_a(x) :=G_a \big|$,
transforms as a gauge field with respect to the local chiral 
rotations generated by $\l (x) := -\frac{{\rm i}}{2}(\s -\bar \s )\big|$.
Here the notation is that $U|:=U (x,\q, \bar \q) \big|_{\q =\bar \q=0}$, 
with $U (x,\q, \bar \q) $ an arbitrary superfield.

Using super-Weyl transformations, one can gauge away $S^{ij}\big|$, 
$Y_{\a\b}\big|$ and some higher-order components of these tensors.
Actually, using both the supergravity gauge transformations and the super-Weyl ones, 
one can choose a Wess-Zumino gauge in which the surviving component fields
match exactly those in the Weyl multiplet  \cite{deWvHVP} except one field 
usually included in 
the Weyl multiplet -- 
the gauge field of  dilatations, $b_m (x)$. However, the latter  is merely a cosmetic 
feature of the superconformal tensor calculus, and has no dynamical impact, 
as it can be algebraically gauged away by local special conformal transformations. 
We hope to discuss these issues in more detail in a separate publication 
in which the supersymmetric action (\ref{InvarAc}) will be reduced to components
in the Wess-Zumino gauge. Actually, there is a simpler independent way 
to justify the claim that the above superspace setting describes the Weyl multiplet.
As will be argued in subsection 2.4, our formulation corresponds to a partial 
gauge fixing in Howe's formulation for $\cN=2$ conformal supergravity \cite{Howe}.
Such a gauge fixing eliminates only  component fields which can algebraically 
be gauged away.   
Therefore, the superspace  setting  presented is adequate to describe the Weyl multiplet.

\subsection{Reduced formulation}
The super-Weyl gauge freedom can be used to gauge away the real or the imaginary 
part of $S^{ij}$. For concreteness, let us choose the first option and impose the gauge
condition
\be 
S^{ij} ={\rm i} {\cS}^{ij}~,  
\qquad {\bar \cS}^{ij} =  \cS^{ij}~.
\ee
Then, the residual super-Weyl transformations are generated 
by a covariantly chiral parameter $\sigma$ constrained as follows:
\bea
\big(\cD^{\a(i}\cD^{j)}_\a +4 {\rm i}\, \cS^{ij} \big)\s=
-\big(\cDB^{(i}_\ad\cDB^{\ad j)} -4{\rm i} \, \cS^{ij} \big){\bar \s}~.
\eea
Such a setting is also  adequate to describe the Weyl multiplet.

\subsection{Comments on  Howe's formulation}

As mentioned earlier, 
the structure group 
in Howe's formulation for $\cN=2$ conformal supergravity \cite{Howe}
is ${\rm SO}(3,1) \times {\rm SU}(2) \times {\rm U}(1)$.
The constraints on the geometry of superspace, 
which were postulated in \cite{Howe}, 
 are invariant under super-Weyl
transformations generated  by a real unconstrained superfield $U$.
The general solution to the constraints involves more irreducible components
for the torsion
than the set given in section 2. The main difference from Grimm's formulation
\cite{Grimm} is that in \cite{Howe} there  occurs an additional tensor
 $G^{ij}_{\a \dot \a}=G^{ji}_{\a \dot \a}$,
along with  the vector $G_{\a \dot \a} $ present in \cite{Grimm}.
The super-Weyl transformations act on $G^{ij}_{\a \dot \a}$ according to
\bea
\d G^{ij}_{\a \dot \a}~ = ~  U \,G^{ij}_{\a \dot \a}
+ c \,[\cD^{(i}_\a , {\bar \cD}^{j)}_{\dot \a} ] U~,
\eea
for some non-zero numerical coefficient $c$.
The constraints are such that $G^{ij}_{\a \dot \a}$ can be gauged away
by super-Weyl transformations.
Then, it can be shown that the U(1) connection 
can   completely be gauged away by 
corresponding U(1)-gauge transformations.
In the gauge $G^{ij}_{\a \dot \a} =0$, the residual super-Weyl freedom 
is described by a parameter constrained by 
\be
 [\cD^{(i}_\a , {\bar \cD}^{j)}_{\dot \a} ] U = 0\quad \Longrightarrow \quad 
 U =\hf ( \s +{\bar \s}) ~, \qquad {\bar \cD}_{\dot \a i} \s =0~.
\label{super-Weyl-gauge}
\ee 
In this super-Weyl gauge,  Howe's formulation reduces to that 
described in section 2.
The action for conformal supergravity in superspace is \cite{Muller}
\bea
S= \int {\rm d}^4x \,{\rm d}^4 \Q \, \cE \,W^{\a\b}W_{\a\b} ~+~{\rm c.c.} ~,
\eea
with $\cE (x, \Q)$ the chiral density\footnote{Here the Grassmann variables $\Q$'s, 
which are used to parametrize covariantly chiral superfields and chiral densities, 
were introduced in \cite{WZ2,Ramirez}, see \cite{Muller} for a review.}, 
is super-Weyl invariant before imposing 
the super-Weyl gauge (\ref{super-Weyl-gauge}).
Therefore, upon fixing this gauge, the action 
remains invariant under the restricted super-Weyl transformations 
(\ref{super-Weyl1}), with $\s$ covariantly chiral. 

The above picture  is completely analogous to the situation in 4D $\cN=1$ supergravity.
To describe the multiplet of conformal supergravity in superspace,
one can introduce a set of constraints that are  invariant under super-Weyl 
transformations generated by a complex unconstrained superfield $ L$
\cite{Gates}  (see \cite{BK} for a review). 
The torsion components are given in terms of a spinor superfield $T_\a$,
chiral superfields $R$ and $W_{(\a \b\g)}$, and a real vector $G_{\a \dot \a}$.
The super-Weyl transformations can be used to gauge away $T_\a$.
In the gauge $T_\a=0$, one stays with a residual super-Weyl invariance 
described by $ L=\hf \s -{\bar \s}$, with $\s$ a covariantly chiral superfield
\cite{HT}.
The resulting formulation, which is known as the old minimal formulation of $\cN=1$ 
supergravity \cite{WZ-s},  is perfectly suited to describe $\cN=1$ conformal supergravity. 
It is much easier to work with  than the original formulation. 

\section{Projective supermultiplets}
\setcounter{equation}{0}

Before introducing an important  family of covariant  multiplets in curved superspace, 
it is worth recalling the definition of rigid projective superfields
\cite{KLR,Lindstrom:1989ne,LR,KLT}. 
\subsection{Review of rigid projective superspace}
In flat global $\cN=2$
superspace  ${\mathbb R}^{4|8}$ parametrized  
by    $ z^A = (x^a,  \q^\a_i, {\bar \q}^i_\ad )$, 
the spinor covariant derivatives obey the algebra:
\bea
\{ D^i_\a, D^j_\b \} = \{ {\bar D}_{\dot \a i} , {\bar D}_{\dot \b j} \}=0~,
\qquad \{ D^i_\a, {\bar D}_{\dot  \b  j } \} =-2{\rm i} \,
\d^i_j (\s^c)_{\a \dot \b}\, \pa_c~.
\eea
Making use of  an isotwistor $u^+_i \in {\mathbb C}^2 \setminus \{0\}$
one may introduce  a subset 
of spinor covariant derivatives
$D^+_\a :=D^i_\a \,u^+_i $ and ${\bar D}^+_{\dot \a} :={\bar D}^i_{\dot \a} \,u^+_i$
that  are linear {\it holomorphic} functions of  $u^+$ and 
strictly anticommute
\bea
\{ D^+_\a , D^+_\b\}& =& \{ {\bar D}^+_{\dot \a} , {\bar D}^+_{\dot \b} \}
=\{ D^+_\a , {\bar D}^+_{\dot \b}\}=0~.
\eea
A projective superfield $Q(z, u^+)$ is defined to obey the constraints  
$D^+_\a Q = {\bar D}^+_{\dot \a} Q=0$
and be a {\it holomorphic  homogeneous} function of $u^+$, 
$Q(z, c\, u^+) =c^n \, Q(z, u^+)$, with $ c\in {\mathbb C}^*:={\mathbb C} \setminus \{0\}$, 
living on an open domain of  ${\mathbb C}^2 \setminus \{0\}$.
Thus, the isotwistor
$u^+_i \in {\mathbb C}^2 \setminus\{0\}$ appears to be  defined modulo 
the equivalence relation $u^+_i \sim c\,u^+_i$,  with $ c\in {\mathbb C}^*$, 
hence the superfields introduced live on
projective superspace  ${\mathbb R}^{4|8} \times {\mathbb C}P^1$.
The projective multiples are holomorphic with respect to a local 
complex coordinate $\z$ used to parametrize  ${\mathbb C}P^1$.
In the north chart of ${\mathbb C}P^1$, where $u^{+\1}\neq 0$, 
this coordinate 
can be defined in the standard way:
$u^{+i} =u^{+\1}(1,\z)$.

\subsection{A projective superspace for supergravity}
We consider curved 4D $\cN=2$ superspace, 
in complete analogy with the 
case of 5D $\cN=1$ supergravity \cite{KT-M2,KT-M3,KT-M4}:
that is, we view the isotwistor  variables $u^{+}_i \in  
{\mathbb C}^2 \setminus  \{0\}$ to be local coordinates that are {\em inert} with respect to 
the subgroup SU(2) of the supergravity gauge group. The reason for doing this is
that it allows us to keep the coordinates $u^{+}_i $ constant; if they transformed
under the local SU(2) gauge symmetry, $\cD^i_{ \a}u^{+}_j$ could not vanish
because of the form of the constraints (\ref{acr1}).
For most applications, it is sufficient to work 
with a large family of the isotwistor superfields, $U^{(n)}(z,u^+)$,
which are described in detail in the appendix and which possess well-defined 
transformation laws with respect to the supergravity gauge group. It is important to note
that since the $u^{+}_i $ are constant, $U^{(n)}(z,u^+)$ is {\em not} a scalar field.
Indeed, {\bf all equations involving $u^{+}_i$ must be homogeneous in $u^{+}_i$ to 
be covariant}. In this approach, the $u^{+}_i$ serve merely to totally symmetrize
all SU(2) indicies.

It might well be interesting to consider a projective superspace formalism where
the $u^{+}_i$ {\em do} transform under the gauge SU(2); in that case, we would have
to find appropriate constraints to avoid introducing new degrees of freedom into the
theory. We leave this for future research.

The operators $\cD^+_{ \a}:=u^+_i\,\cD^i_{ \a} $ and   
${\bar \cD}^+_{\dot  \a}:=u^+_i\,{\bar \cD}^i_{\dot \a} $
map the isotwistor superfields into isotwistor superfields with $\frac12$ unit
higher isospin
and obey the (isospin 1) anti-commutation relations:
\bea
\{  \cD^+_{ \a} , \cD^+_{ \b} \}
=4\, Y_{\a \b}\,J^{++}
+4 \, S^{++}M_{\a \b}~, \qquad 
\{\cD_\a^+,\cDB_\bd^+\}=
8 \,G_{\a \bd} \,J^{++}~,
\label{analyt}
\eea
where 
$J^{++}:=u^+_i u^+_j \,J^{ij}$ and 
$S^{++}:=u^+_i u^+_j \,S^{ij}$. 

A projective supermultiplet of weight $n$,
$Q^{(n)}(z,u^+)$, is a constrained isotwistor superfield. 
Specifically, it 
is a scalar superfield that 
lives on  $\cM^{4|8}$, 
is holomorphic with respect to 
the isotwistor variables $u^{+}_i $ on an open domain of 
${\mathbb C}^2 \setminus  \{0\}$, 
and is characterized by the following conditions:\\
(i) it obeys the covariant analyticity constraints 
\be
\cD^+_{\a} Q^{(n)}  = {\bar \cD}^+_{\ad} Q^{(n)}  =0~;
\label{ana}
\ee  
(ii) it is  a homogeneous function of $u^+$ 
of degree $n$, that is,  
\be
Q^{(n)}(z,c\,u^+)\,=\,c^n\,Q^{(n)}(z,u^+)~, \qquad c\in \mathbb{C}^*~;
\label{weight}
\ee
(iii)  gauge transformations (\ref{tau}) act on $Q^{(n)}$ 
as follows:
\bea
\d_K Q^{(n)} 
&=& \Big( K^{{C}} \cD_{{C}} + K^{ij} J_{ij} \Big)Q^{(n)} ~,  
\non \\ 
K^{ij} J_{ij}  Q^{(n)}&=& -\frac{1}{(u^+u^-)} \Big(K^{++} D^{--} 
-n \, K^{+-}\Big) Q^{(n)} ~, \qquad 
K^{\pm \pm } =K^{ij}\, u^{\pm}_i u^{\pm}_j ~,
\label{harmult1}   
\eea 
where
\bea
D^{--}=u^{-i}\frac{\partial}{\partial u^{+ i}} ~,\qquad
D^{++}=u^{+ i}\frac{\partial}{\partial u^{- i}} ~.
\label{5}
\eea
The transformation law (\ref{harmult1}) involves an additional isotwistor,  $u^-_i$, 
which is subject 
to the only condition $(u^+u^-) := u^{+i}u^-_i \neq 0$, and is otherwise completely arbitrary.
By construction, $Q^{(n)}$ is independent of $u^-$, 
i.e. $\pa  Q^{(n)} / \pa u^{-i} =0$,
and hence $D^{++}Q^{(n)}=0$.
One can see that $\d_K Q^{(n)} $ 
is also independent of the isotwistor $u^-$, $\pa (\d_K Q^{(n)})/\pa u^{-i} =0$,
due to (\ref{weight}). 
It follows  from (\ref{harmult1})
\bea
J^{++} \,Q^{(n)}=0~, \qquad J^{++} \propto D^{++}~,
\label{J++}
\eea
and hence the covariant analyticity constraints (\ref{ana}) are indeed consistent.

It follows from (\ref{S-analit}) that 
\be
S^{++}:= S^{ij}u^+_iu^+_j~, 
\qquad 
\widetilde{S}^{++}:= {\bar S}^{ij} u^+_iu^+_j
\ee
are projective supermultiplets of weight $+2$, 
\be
\cD^+_{\a} S^{++}  = {\bar \cD}^+_{\ad} S^{++}  =0~.
\ee  

Let $Q^{(n)} (z,u^+) $ be a projective supermultiplet of weight $n$.
Assuming that it transforms homogeneously under the super-Weyl transformations,
the analyticity constraints uniquely fix its transformation law:
\be
\d_{\s} Q^{(n)} =\frac{n}{2} \big(\s +\bar \s \big)  Q^{(n)} ~.
\label{Q(n)super-Weyl}
\ee
The assumption of homogeneity of the transformation law is crucial for 
the derivation of (\ref{Q(n)super-Weyl}); there are some fields, such as 
the torsion component $S^{++}$, which is a projective multiplet of weight $+2$,
that transform inhomogeneously under the super-Weyl transformation,
\be
\d_{\s} S^{++} ={\bar \s} S^{++} -\frac{1}{4}(\cD^+)^2 \s~,
\ee
in accordance with (\ref{super-Weyl-S}).

Given a projective multiplet $Q^{(n)}(z,u^+)$,
its complex conjugate 
is not covariantly analytic.
However, similar to the case of flat superspace, 
one can introduce a generalized,  analyticity-preserving 
conjugation, $Q^{(n)} \to \widetilde{Q}^{(n)}$, defined as
\be
\widetilde{Q}^{(n)} (u^+)\equiv \bar{Q}^{(n)}\big(
\overline{u^+} \to 
\widetilde{u}^+\big)~, 
\qquad \widetilde{u}^+ = {\rm i}\, \s_2\, u^+~, 
\ee
with $\bar{Q}^{(n)}(\overline{u^+}) $ the complex conjugate of $Q^{(n)}$.
It is not difficult to check that $\widetilde{Q}^{ (n) } (z,u^+)$ is a projective multiplet of weight $n$.
One can see that
$\widetilde{\widetilde{Q}}{}^{(n)}=(-1)^nQ^{(n)}$,
and therefore real supermultiplets can be consistently defined when 
$n$ is even.
In what follows, $\widetilde{Q}^{(n)}$ will be called the smile-conjugate of 
${Q}^{(n)}$. Geometrically, this conjugation is complex conjugation composed
with the antipodal map on the projective space ${\mathbb C}P^1$.

Consider a supergravity background. The superconformal 
group of this space is defined to be generated by those combined
infinitesimal transformations  (\ref{tau}) and (\ref{super-Weyl1})
which do not change the covariant derivatives,
\be
\d_K \cD_{A} + \d_\s \cD_{A} =0~.
\ee
This definition is analogous to that often used in 4D $\cN=1$ supergravity
\cite{BK}.
In the case of 4D $\cN=2$ flat superspace, it is equivalent to the definition 
of the superconformal Killing vectors, see  \cite{K2} and references therein. 
In this case, the transformation laws of the projective multiplets reduce to those describing 
the rigid superconformal projective multiplets \cite{K2}.

To gain further insight into the structure of projective supermultiplets 
$Q^{(n)}(z,u^+)$, it is instructive to switch from their description in terms
of the homogeneous coordinates, $u^+_i$, for ${\mathbb C}P^1$
to a formulation that makes use of the inhomogeneous local complex 
variable $\z$ which is invariant under the projective rescalings $u^+_i\to cu^+_i$.
In such a setting, one should replace $Q^{(n)}(z,u^+)$ with a new superfield 
$Q^{[n]}(z,\z) \propto Q^{(n)}(z,u^+)$, where $Q^{[n]}(z,\z) $ is  holomorphic 
with respect to  $\z$, and its explicit definition depends on the supermultiplet under 
consideration.  
One can cover  ${\mathbb C}P^1$  
by two open charts in which $\z$ can be defined, 
and the simplest choice is:
(i) the north chart characterized by $u^{+\1}\neq 0$;
(ii) the south chart with  $u^{+\2}\neq 0$.
Below, our consideration  will be restricted to the north chart.

In the north chart $u^{+\1}\neq 0$, 
the projective-invariant  variable $\z \in \mathbb C$ is defined as
\bea
u^{+i} =u^{+\1}(1,\z) =u^{+\1}\z^i ~,\qquad 
\z^i=(1,\z)~, \qquad \z_i= \ve_{ij} \,\z^j=(-\z,1)~.
\label{zeta}
\eea
Since any projective multiplet $Q^{(n)}$ and its
transformation (\ref{harmult1}) do not depend on $u^-$, 
we can make a convenient choice for the latter.
In the north chart, it is
\be
u^-_i =(1,0) ~, \qquad   \quad ~u^{-i}=\ve^{ij }\,u^-_j=(0,-1)~.
\label{fix-u-}
\ee   
The transformation parameters $K^{++}$  
and $K^{+-}$ in (\ref{harmult1}) 
can be represented as $K^{++} =\big(u^{+\1}\big)^2 {K}^{++} (\z)$ 
and $K^{+-}= u^{+\1}K(\z) $, where
\bea
{K}^{++} (\z)&=& {K}^{ij} \,\z_i \z_j
=  {K}^{\1 \1 }\, \z^2 -2  {K}^{\1 \2}\, \z 
+ {K}^{\2 \2} ~,
\quad
K(\z)= {K}^{\1 i} \,\z_i 
=  - {K}^{\1 \1} \,\z + {K}^{\1 \2}  ~.~~~~~~~
\label{K++K}
\eea
If the projective supermultiplet  $Q^{(n)}(z,u^+)$ 
is described by $Q^{[n]}(z,\z) \propto Q^{(n)}(z,u^+)$ in the north chart, 
then the covariant analyticity conditions (\ref{ana}) becomes
\bea
\cD^+_{ \a} (\z) \, Q^{[n]} (\z) &=&0~, 
\qquad \cD^+_{ \a} (\z) = -\cD^i_{ \a} \z_i
=\z \,\cD^{\1}_{ \a}  - \cD^{\2}_{ \a} ~, \non \\
{\bar \cD}^{+ \ad} (\z) \, Q^{[n]} (\z) &=&0~, 
\qquad {\bar \cD}^{+ \ad} (\z) 
= {\bar \cD}^\ad_i \z^i = {\bar \cD}^\ad_{\1}  +\z{\bar \cD}^\ad_{\2}~.
\label{ana2}
\eea
Let us give several important examples of projective supermultiplets.

An {\it arctic} multiplet\footnote{We follow the terminology introduced
in the rigid supersymmetric case in  \cite{G-RLRvUW}.}
 of weight $n $ is defined to be holomorphic 
on the north chart. It can be represented as 
\bea
\U^{(n)} (z, u) =  (u^{+\1})^n\, \U^{[n]} (z, \z) ~, \quad \qquad
\U^{ [n] } (z, \z) = \sum_{k=0}^{\infty} \U_k (z) \z^k~.
\label{arctic1}
\eea
The transformation law of $\U^{[n]}$ can be read off from 
eq. (\ref{harmult1}) by noting 
(see \cite{K,K2} for technical  details)
\bea
K^{ij}J_{ij}\, \U^{ [n] } (\z)
 =  \Big(  {K}^{++} (\z)\,\pa_\z + n\,K (\z) \Big) \U^{[n]}(\z)~,
\label{arctic2}
\eea
or equivalently
\bea
J_{\1 \1} \U_0 =0~, \qquad 
J_{\1 \1} \U_k &=&(k-1-n) \U_{k-1}~, \quad k>0 
\non \\
J_{\2 \2} \U_k &=&(k+1) \U_{k+1}~,  
\label{arctic3} \\
J_{\1 \2} \U_k &=&(\frac{n}{2}-k) \U_{k}~. \non 
\eea
Eq. (\ref{arctic3}) defines an  infinite dimensional representation of 
the Lie algebra su(2). 
It should be  emphasized  that the transformation  of  $\U^{[n]}$ 
preserves the functional structure of  $\U^{[n]}$  defined in (\ref{arctic1}).

The constraints (\ref{ana2}) imply
\bea
 {\bar \cD}^\ad_{\1}\U_0 &=& 0~, 
 \qquad 
  {\bar \cD}^\ad_{\1} \U_1 = -  {\bar \cD}^\ad_{\2} \U_0~, \non \\
\cD^{\2}_{\a}  \U_0&=&0~, 
\qquad 
\cD^{\2}_{ \a}  \U_1 =\cD^{\1}_{ \a}  \U_0~.
\label{ana-arctic}
\eea
The integrability conditions for these constraints can be shown to be 
$J_{\1 \1} \U_0 = 0$ and $J_{\1 \1} \U_1 =- 2 J_{\1 \2} \U_0$, and they 
hold identically due to (\ref{arctic3}).
Using the anticommutation relations (\ref{acr1}) and (\ref{acr2}),
one can deduce from (\ref{ana-arctic}) 
\bea
-\frac{1}{4} \Big[ 
({\bar \cD}_{\1})^2 + 4 {\bar S}^{\2 \2} \Big] \U_1 
&=& n \,{\bar S}^{\1 \2}  \U_0~, \non \\
-\frac{1}{4} \Big[ 
( \cD^{\2})^2 + 4 S^{\2 \2} \Big] \U_1 
&=& n \,S^{\1 \2}  \U_0~.
\eea

The smile-conjugate of $ \U^{(n)}$ will be called 
an {\it antarctic} multiplet of weight $n $. It proves to be  holomorphic on the south
chart, while  in the north chart it has the form
\bea
\widetilde{\U}^{(n)} (z, u) &=&  (u^{+\2})^n\, \widetilde{\U}^{[n]} (z, \z)~, \qquad
\widetilde{\U}^{[n]} (z, \z) = \sum_{k=0}^{\infty} (-1)^k {\bar \U}_k (z)
\frac{1}{\z^k}~,
\label{antarctic1}
\eea
with $ {\bar \U}_k$ the complex conjugate of $\U_k$.
Its  transformation follows from (\ref{harmult1}) by noting
\bea
 K^{ij}J_{ij}\,  \widetilde{\U}^{[n]} (\z)=  
 \frac{1}{\z^n}\Big(  {K}^{++} (\z) \,\pa_\z 
 + n\,K (\z) 
 \Big) \Big(\z^n\,\widetilde{\U}^{(n)} (\z)\Big)~.
\label{antarctic2}
\eea
The arctic multiplet $ \U^{[n]} $ and its smile-conjugate $\widetilde{\U}^{(n)} $ 
constitute a polar multiplet.

The simplest  projective supermultiplets are
real $O(2n)$-multiplet, with $n=1,2,\dots$ 
\bea
H^{(2n)}(z,u^+)&=&u^+_{i_1}\cdots u^+_{i_{2n}}\,H^{i_1\cdots i_{2n}}(z)~, 
\qquad \widetilde{H}^{(2n)} =H^{(2n)}~.
\eea
Here the case $n=1$ corresponds to the $\cN=2$ tensor multiplet
\cite{N=2tensor,BS}.
Such multiplets are holomorphic on ${\mathbb C}P^1$.
We can represent 
\bea
H^{(2n)}(z,u^+)&=&
\big({\rm i}\, u^{+\1} u^{+\2}\big)^n H^{[2n]}(z,\z) ~, 
\non \\
H^{[2n]}(z,\z) &=&
\sum_{k=-n}^{n} H_k (z) \z^k~,
\qquad  {\bar H}_k = (-1)^k H_{-k} ~. 
\label{o2n1}
\eea
The transformation  of $H^{[2n]} $ follows from (\ref{harmult1}) by noting
\bea
 K^{ij}J_{ij}\,  H^{[2n]} &=&  
 \frac{1}{\z^n}\Big(    {K}^{++} (\z) \,\pa_\z +2n\,K(\z)
 \Big) \Big(\z^n H^{[2n]} \Big)~.
\label{o2n2}
\eea
This can be seen to be equivalent to 
\bea
J_{\1 \1} H_{-n} =0~, \qquad 
J_{\1 \1} H_k &=&(k-1-n) H_{k-1}~, \quad  -n<k \leq n 
\non \\
J_{\2 \2} H_{n} =0~, \qquad
J_{\2 \2} H_k &=&(k+1+n) H_{k+1}~,  \quad  -n \leq k < n 
\label{o2n3} \\
J_{\1 \2} H_k &=& -k H_{k}~. \non 
\eea
The constraints (\ref{ana2}) imply
\bea
 {\bar \cD}^\ad_{\1}H_{-n} &=& 0~, 
 \qquad 
  {\bar \cD}^\ad_{\1} H_{-n+1} = -  {\bar \cD}^\ad_{\2} H_{-n}~, \non \\
\cD^{\2}_{\a}  H_{-n}&=&0~, 
\qquad 
\cD^{\2}_{ \a}  H_{-n+1} =\cD^{\1}_{ \a}  H_{-n}~,
\label{o2n4}
\eea
and from here one deduces
\bea
-\frac{1}{4} \Big[ 
({\bar \cD}_{\1})^2 + 4 {\bar S}^{\2 \2} \Big] H_{-n+1}
&=& 2n \,{\bar S}^{\1 \2}  H_{-n}~, \non \\
-\frac{1}{4} \Big[ 
( \cD^{\2})^2 + 4 S^{\2 \2} \Big] H_{-n+1} 
&=& 2n \,S^{\1 \2}  H_{-n} ~.
\eea

Another important projective multiplet 
is a real {\it tropical} multiplet of weight $2n$:
\bea
U^{(2n)} (z,u^+) &=& 
\big({\rm i}\, u^{+\1} u^{+\2}\big)^n U^{[2n]}(z,\z) =
\big(u^{+\1}\big)^{2n} \big({\rm i}\, \z\big)^n U^{[2n]}(z,\z)~, \non \\
U^{[2n]}(z,\z) &=&
\sum_{k=-\infty}^{\infty} U_k (z) \z^k~,
\qquad  {\bar U}_k = (-1)^k U_{-k} ~. 
\label{trop-nj}
\eea
The SU(2)-transformation law of $U^{[2n]}(z,\z) $
 copies (\ref{o2n2}). To describe 
 a massless vector multiplet prepotential, one should choose $n=0$.
 Supersymmetric real Lagrangians correspond to the choice $n=1$, see below.

\section{Coupling to  vector supermultiplets}
\setcounter{equation}{0}

The  multiplet of conformal supergravity 
can naturally be coupled to off-shell vector multiplets.
Let us describe in detail the case 
of a single Abelian vector multiplet, due to its importance 
for the subsequent consideration.\footnote{An extension to the non-Abelian case is not difficult.}
Its coupling to the Weyl multiplet
is achieved, first of all,  by modifying the covariant derivatives as follows:
\bea
\cD_{{A}} \quad \longrightarrow \quad 
{\bm \cD}_{{A}}:=\cD_{{A}}+ V_{{A}}{\bm Z}~,
\eea
with $V_{{A}}(z) $ the gauge connection.
It is convenient to interpret the generator $\bm Z $ as  a real central charge.
In addition, one should impose appropriate covariant constraints, 
guided by the rigid supersymmetric formulation for the vector multiplet \cite{GSW},
on some components of the gauge-invariant field strength $F_{AB}$  which appears
in the algebra of gauge-covariant derivatives 
\bea
{[}{\bm \cD}_{{A}},{\bm \cD}_{{B}}\}&=&
T_{{A}{B}}{}^{{C}}\, {\bm \cD}_{{C}}
+\hf R_{{A}{B}}{}^{cd}M_{c d}
+R_{{A}{B}}{}^{kl}J_{kl}
+F_{{A} {B}}{\bm Z}~.
\eea
Here the torsion and curvature are the same as 
in eq. (\ref{algebra}).

The  components of $F_{AB}$  are:
\begin{subequations}
\bea
F_{\a}^i{}_\b^j&=&-2\ve_{\a\b}\ve^{ij}\bar{W}~,\qquad
F^{\ad}_i{}^\bd_j=2\ve^{\ad\bd}\ve_{ij}{W}~,\qquad
F_{\a}^i{}^\bd_j=0~,  \\
F_{a}{}_{\b}^j&=&
{\frac{\ri}2}(\s_a)_\b{}^{\gd}\cDB_\gd^j\bar{W}~,\qquad 
F_{a}{}^{\bd}_j=
-{\frac{\ri}2}(\s_a)_\g{}^{\bd}\cD^\g_j{W}~, \\
F_{ab}&=&
-{\frac18}(\s_{ab})_{\b\g}\cD^{\b k}\cD^{\g}_k{W}
-{\frac18}(\ts_{ab})_{\bd\gd}\cDB^{\bd k}\cDB^{\gd}_k\bar{W}
\non\\
&&
+{\frac12}\Big(({\ts}_{ab})_{\ad\bd}\bar{W}^{\ad\bd}
-({\s}_{ab})_{\a\b}Y^{\a\b}\Big)W
-{\frac12}\Big(({\s}_{ab})_{\a\b}{W}^{\a\b}
-({\ts}_{ab})_{\ad\bd} \bar{Y}^{\ad\bd}\Big)\bar{W} 
~.~~~~~~
\eea
\end{subequations}
Here $W$ is a covariantly chiral superfield,
\be
{\bar \cD}_{\ad i} W =0~,
\ee
obeying  the Bianchi identity
\bea
\Big(\cD^{\g(i}\cD_\g^{j)}+4S^{ij}\Big)W
&=&
\Big(\cDB_\gd^{(i}\cDB^{ j) \gd}+ 4\bar{S}^{ij}\Big)\bar{W}
~.
\label{vectromul}
\eea

Under the super-Weyl transformations,   $W$ varies as 
\be
\d_{\s} W = \s W~.
\label{Wsuper-Weyl}
\ee
Introduce
\be
\S^{++}:=\frac{1}{4}\Big( (\cD^+)^2 +4S^{++}\Big)W
=\S^{ij}u^+_i u^+_j~.
\label{Sigma}
\ee
Using (\ref{vectromul}), one can show that $\S^{++}$ is a real 
projective supermultiplet of weight $+2$, 
\be
\cD^+_{\a} \S^{++}  = {\bar \cD}^+_{\ad} \S^{++}  =0~, \qquad
\widetilde{\S}^{++}=\S^{++}~.
\label{Sigma-con}
\ee  
The super-Weyl transformation of $\S^{++}$
is 
\be
\d_{\s} \S^{++} = \big(\s +\bar \s \big)  \S^{++} ~,
\label{S(++)super-Weyl}
\ee
compare with (\ref{Q(n)super-Weyl}). 

The super-Weyl gauge freedom can be used to choose the gauge 
\be
W=-{\rm i}~,
\label{W=-i}
\ee
which is the flat-superspace value of the rigid central charge, 
see \cite{DIKST} for a related discussion.
In this gauge, eq. (\ref{vectromul}) reduces to 
\bea
S^{++}={\rm i} \cS^{++} ~, \qquad 
\cS^{++}= \widetilde{ \cS}^{++}~,
\eea
with $\cS^{++}$ a real $O(2)$
multiplet. 
As a result, one arrives at 
the well-known superspace realization \cite{Howe,Muller} for 
the minimal multiplet for $\cN=2$ supergravity \cite{BS}.

Consider now a system of several Abelian vector multiplets, and let $W^\m$ be 
their covariantly chiral field strengths. Let $F(W^\m)$ be a holomophic homogeneous function 
of degree one, $F(c\,W^\m)= \,cF(W^\m)$. Then, we can define a generalization of $ \S^{++}$
(\ref{Sigma}): 
\be
{\bm \S}^{++}:=\frac{1}{4}\Big( (\cD^+)^2 +4S^{++}\Big)F(W^\m)
={\bm \S}^{ij}u^+_i u^+_j~, 
\qquad F(c\,W^\m)= \,cF(W^\m)~.
\label{Sigma2}
\ee
This superfield is not real, ${\bm \S}^{++} \neq \widetilde{{\bm \S}}{}^{++}$,
If $F$ is not linear. 
However, it enjoys the other properties of $\S^{++}$ given in eqs. (\ref{Sigma-con})
and (\ref{S(++)super-Weyl}). 

\section{Action principle}
\setcounter{equation}{0}

Let $\cL^{++}$ be a real projective multiplet of weight two.
In particular, its super-Weyl transformation is
\be
\d_{\s} \cL^{++} = \big(\s +\bar \s \big)  \cL^{++} ~.
\label{L(++)super-Weyl}
\ee 
Associated with $\cL^{++}$ is the following functional 
\bea
S&=&
\frac{1}{2\pi} \oint (u^+ \rd u^{+})
\int \rd^4 x \,{\rm d}^8\q\,E\, \frac{W{\bar W}\cL^{++}}{(\S^{++})^2}~, 
\qquad E^{-1}= {\rm Ber}(E_A{}^M)~.
\label{InvarAc}
\eea
This functional is  obviously invariant under  re-scalings
$u_i^+(t)  \to c(t) \,u^+_i(t) $, for an arbitrary function
$ c(t) \in {\mathbb C}\setminus  \{0\}$, 
where $t$ denotes the evolution parameter 
along the closed integration contour.
Since $E$ is invariant under the super-Weyl transformations,
\be
\d_{\s} E=0~,
\ee
eqs. (\ref{Wsuper-Weyl}), (\ref{S(++)super-Weyl}) and (\ref{L(++)super-Weyl})
show that $S$ is super-Weyl invariant.
The action can also be shown to be invariant under arbitrary supergravity gauge 
transformations, in complete analogy with the 5D considerations of 
\cite{KT-M3,KT-M4}.

One can represent $\cL^{++}$ in the form
\bea
\cL^{++}(z,u^+)&=&\frac{1}{16}\Big( ({\bar \cD}^+)^2 +4\widetilde{S}^{++}\Big) 
\Big( (\cD^+)^2 +4S^{++}\Big)\cU^{(-2)}(z,u^+)\non \\
&=&\frac{1}{16}\Big( (\cD^+)^2 +4S^{++}\Big) \Big( ({\bar \cD}^+)^2 +4\widetilde{S}^{++}\Big) 
\cU^{(-2)}(z,u^+)~, \
\eea
for some projective prepotential $\cU^{(-2)}$ 
which is an example of the isotwistor superfields introduced 
in the appendix. 
It can be seen that $\cU^{(-2)}$ should be inert under the super-Weyl 
transformations, 
\be 
\d_\s \cU^{(-2)} =0~,
\ee
 in order for $\cL^{++} $ to possess the transformation law 
(\ref{L(++)super-Weyl}).
Then, the action (\ref{InvarAc}) can be rewritten as
\be
S= \frac{1}{2\pi} \oint (u^+ \rd u^{+})
\int \rd^4 x \,{\rm d}^8\q\,E\, \cU^{(-2)}~.
\label{InvarAcU}
\ee
This relation leads to the following  important result:   
if $\cL^{++}$ is generated in terms of some supermultiplets
to which  the central charge vector multiplet does not belong, 
then the action  $S$ is independent of  the vector multiplet chosen.

Let us demonstrate that in a flat superspace limit, eq. (\ref{InvarAcU}) is equivalent 
to the action principle in projective superspace \cite{KLR}. 
Let $D_{{A}} =(\pa_{{a}}, D_{{\a}}^i,{\bar D}^\ad_i)$ be the flat covariant 
derivatives. We also denote by $L^{++} $ and $U^{(-2)}$ 
the flat-superspace limits of $\cL^{++} $ and $\cU^{(-2)}$, 
\bea
L^{++}(z,u^+) =(D^+)^4 U^{(-2)}~, \qquad 
(D^+)^4=\frac{1}{16} ({\bar D}^+)^2 (D^+)^2
&=&\frac{1}{16} (D^+)^2  ({\bar D}^+)^2~.
\eea
The flat-superspace version of (\ref{InvarAcU}),  
\be
S_{\rm flat}= \frac{1}{2\pi} \oint (u^+ \rd u^{+})
\int \rd^4 x \,{\rm d}^8\q\, U^{(-2)}~,
\label{InvarAcU-flat}
\ee
can equivalently be rewritten as 
\bea
S_{\rm flat}&=&
\frac{1}{2\pi} \oint \frac{(u^+\rd u^{+})}{(u^+u^-)^4}
\int \rd^4 x \, (D^-)^4(D^+)^4U^{(-2)} \Big| 
\non \\
&=& \frac{1}{2\pi}  \oint \frac{(u^+\rd u^{+})}{(u^+u^-)^4}
\int \rd^4 x \, (D^-)^4 L^{++}\Big|~,
\label{flatac}
\eea
where the spinor derivatives $D^-_\a$ and ${\bar D}^-_\ad$ 
are obtained from $D^+_\a$ and ${\bar D}^+_\ad$ by replacing 
$u^+_i \to u^-_i$, with the latter fixed (i.e. $t$-independent)
isotwistor obeying the only constraint 
$(u^+(t)u^-)\neq 0$ at each point of  the integration contour. 
This is exactly the projective superspace action \cite{KLR}
as reformulated in \cite{Siegel}.
The action can be seen to be invariant 
under arbitrary projective transformations of the form:
\be
(u_i{}^-\,,\,u_i{}^+)~\to~(u_i{}^-\,,\, u_i{}^+ )\,R~,~~~~~~R\,=\,
\left(\begin{array}{cc}a~&0\\ b~&c~\end{array}\right)\,\in\,{\rm GL(2,\mathbb{C})}~.
\label{projectiveGaugeVar}
\ee
Without loss of generality, we can assume the north pole of ${\mathbb C}P^1$ 
is outside of the integration contour, hence $u^+$
can be represented as in  eq. (\ref{zeta}),
with $\z$ the local complex coordinate for   ${\mathbb C}P^1$.
Using the projective invariance (\ref{projectiveGaugeVar}), we can then choose
$u^-_i$ in the form  (\ref{fix-u-}).
${}$Finally, representing $L^{++}$ in the form
\be
L^{++}(z,u^+) =  {\rm i}\, u^{+\1} u^{+\2}\,
L(z,\z) =  {\rm i} \big( u^{+\1} \big)^2 \z\, L(z,\z)~, 
\ee
and also using the fact that $L^{++}$ enjoys the constraints
$\z_i D^i_\a L=\z_i {\bar D}^i_\ad L=0$, we can finally rewrite  
$S_{\rm flat}$ as an integral over the $\cN=1$ superspace
parametrized by the following coordinates: $(x^a, \q^\a_{\1}, {\bar \q}^{\1}_\ad)$. 
The result is
\bea
S_{\rm flat}=
\frac{1}{2\pi \rm i}  \oint \frac{\rd \z}{\z}
\int \rd^4 x \, {\rm d}^4\q\,  L\Big|_{\q_{\2}={\bar \q}^{\2} =0}~.
\eea
This is equivalent to the original form of the projective superspace action \cite{KLR}.

It should be pointed out that the super-Weyl gauge freedom can be 
fixed as in (\ref{W=-i}).
Then, the action (\ref{InvarAc}) becomes 
\bea
S&=&
\frac{1}{2\pi} \oint (u^+ \rd u^{+})
\int \rd^4 x \,{\rm d}^8\q\,E\, \frac{\cL^{++}}{(\cS^{++})^2}~. 
\label{InvarAc2}
\eea
This result can be compared with the 5D $\cN=1$ supergravity action principle 
\cite{KT-M3}.

The approach developed in this paper is well-suited for the off-shell
description of $\cN=2$ Poincar\'e supergravity and its matter couplings. 
Such a description only requires  re-casting the conceptual framework 
of the $\cN=2$ superconformal tensor calculus  
(see \cite{deWvHVP,deWLVP} and references therein) in our 
superspace  setting.
One should consider super-Weyl invariant couplings of 
the Weyl multiplet to supersymmetric matter, and then break 
the super-Weyl invariance. As is known, the set of matter supermultiplets
should include two (conformal) compensators. One of them is universal
and can  be identified with the central charge vector multiplet.  
However, the choice of a second compensator is not unique.
It can be taken to be a hypermultiplet,  or a tensor multiplet, or a nonlinear multiplet.
For concreteness, here we choose the first option.
It is known that the action for the central charge vector multiplet
can be written as a chiral integral \cite{Muller}:
\bea
S= \frac{1}{\k^2} \int {\rm d}^4x \,{\rm d}^4 \Q \, \cE \,W^2  ~+~{\rm c.c.}~,
\eea
with $\cE$ the chiral density, and $\k$ the gravitational coupling constant.
It turns out that this functional can be rewritten in the form (\ref{InvarAc}). 
To achieve this, we should introduce  the gauge field 
of  the central charge vector multiplet, ${\mathbb V}(z,u^+)$, 
which is a real projective weight-zero superfield  (tropical multiplet). 
Then, $\cL^{++}   \propto {\mathbb V} \,\S^{++}$ and 
\bea
\int {\rm d}^4x \,{\rm d}^4 \Q \, \cE \,W^2  \propto 
\frac{1}{2\pi} \oint (u^+ \rd u^{+})
\int \rd^4 x \,{\rm d}^8\q\,E\, \frac{W{\bar W}}{\S^{++}}{\mathbb V}~.
\eea

Now, let us couple the Weyl multiplet to (i) a system of Abelian 
vectors multiplets (including the central charge vector multiplet), 
with $W^\m$ the corresponding covariantly chiral field strengths); 
and (ii) a system of hypermultiplets described 
by weight-one  covariantly arctic multiplets $\U^+(z,u^+)$  and their conjugates
$ \widetilde{\U}^+$'s (defined in complete analogy with  the 5D case \cite{KT-M2}).
The supergravity-matter Lagrangian  can be chosen to be
\be
\cL^{++}   = {\mathbb V} \Big({\bm  \S}^{++} + \widetilde{{\bm \S}}{}^{++}\Big)
-  {\rm i} \, K(\U^+, \widetilde{\U}^+)~,
\label{SUGRA-matter}
\ee
with ${\bm  \S}^{++}$ defined in (\ref{Sigma2}), and 
the real function $K(\F, \bar \F) $
obeying 
the homogeneity condition 
\bea
 \F^I \frac{\pa}{\pa \F^I} K(\F, \bar \F) =  K( \F,   \bar \F)~.
 \label{Kkahler2}
 \eea
 The action possesses the gauge invariance 
\be
\d {\mathbb V} =  \l +\tilde{\l}~, 
\ee
with $\l$ a weight-zero arctic multiplet.
Although this invariance is not obvious, 
it can be established  choosing 
a supergravity Wess-Zumino gauge and applying considerations similar to those given 
in the five-dimensional case  \cite{KT-M2}).

The hypermultiplet sector of (\ref{SUGRA-matter}) is a curved-space 
extension of the rigid superconformal sigma model \cite{K2} 
(a special family of the general $\cN=2$ supersymmetric
nonlinear sigma model \cite{LR}).  
Let $ {\bm \U}^+$ be the compensator contained in our system of covariantly arctic 
multiplets $\U^+$.  By analogy with the flat case \cite{K2},
we can introduce new hypermultiplet variables comprising the unique 
weight-one multiplet ${\bm \U}^+(z,u^+)$ and some set of weight-zero covariantly arctic multiplets
$\u^I(z,u^+)$. We can represent 
\bea
K(\U^+, \widetilde{\U}^+)
= \widetilde{{\bm \U}}{}^+ {\bm \U}^+ \,{\rm e}^{-\cK(\u, \widetilde{\u} )}~,
\eea
with $\cK(\u, \widetilde{\u} )$ a K\"ahler potential. 
This Lagrangian is invariant under K\"ahler tansformations 
\bea
{\bm \U}^+ ~\longrightarrow ~{\rm e}^{\L(\u)} \, {\bm \U}^+~, \qquad 
\cK(\u, \widetilde{\u} ) ~\to ~ \cK(\u, \widetilde{\u} )+ \L(\u) + {\bar \L}( \widetilde{\u})~,
\eea
with $\L $ a holomorphic function. Note that this is precisely the structure uncovered in
\cite{de Wit:2001dj} by considering the geometry of $\cN=2$ supersymmetric
nonlinear sigma models. The potential $K(\U^+, \widetilde{\U}^+)$ has the interpretation of
the hyperk\"ahler potential on the hyperk\"ahler cone, and $\cK(\u, \widetilde{\u} )$ is the K\"ahler potential of the twistor space of the underlying Quaternion K\"ahler geometry. 

\section{Conclusions}
\setcounter{equation}{0}

In this paper we have constructed $\cN=2$ four dimensional (conformal) supergravity in projective superspace. 

Our starting point is the observation that  Grimm's formulation of the superspace constraints and their solutions allow additional Weyl transformations as a symmetry. These enable us to identify the Weyl multiplet residing in Grimm's solution by going to a  Wess-Zumino gauge. Equivalently, our formulation represents a partial gauge fixing of Howe's formulation of $\cN=2$ supergravity.

The transition to projective superspace proceeds via the introduction of isotwistor variables $u^+_i$ in parallel to the rigid case. An important ingredient  is that these are taken to be covariantly constant, a feature which may seem at variance with covariance of the $D$-algebra for $u^+_iD^i_\alpha$. However, we demonstrate explicitly that covariance is maintained when acting on isotwistor superfields. An open question for future investigations is to find a formulation where the relation between the supergravity SU(2) and the isotwistor transformations is carried by a geometric field.

Within our local projective approach we construct various matter couplings as well as 
a superspace action. The restriction to Poincar\'e supergravity is discussed.

Among the future extensions of this work we mention the derivation of the explicit 
$\cN=1$ as well as $\cN=0$ component content of our isotwistor superfields. In particular, it should be possible to compare the Poincar\'e supergravity content to the $\cN=1$ formulation given in
\cite{Gates:1984wc}.

As mentioned in the introduction, it is important for the quantum theory to have a manifest off-shell formulation, and we expect that our results will find applications there. 

Finally, we observed that the geometric structure of hypermultiplets coupled to supergravity
described in \cite{de Wit:2001dj} arises completely naturally.

\noindent
{\bf Acknowledgements:}\\
The work of SMK and GT-M is supported  in part
by the Australian Research Council.
UL acknowledges support by EU grant (Superstring theory) 
MRTN-2004-512194 and by VR grant 621-2006-3365. MR is supported in
part by NSF grant no. PHY-06-53342. MR and SMK thank the 2007 Simons
Workshop in Mathematics and Physics for providing an opportunity to discuss
ideas that contributed to this work. SMK, UL and MR also thank the Workshop in Geometry and Supersymmetry (Uppsala University, November 2007) where further discussions 
toward this project took place.

\begin{appendix}

\section{Isotwistor superfields}
\setcounter{equation}{0}

Consider a completely symmetric isotensor superfield, 
$F^{i_1 \dots i_n}(z) = F^{(i_1 \dots i_n)}(z)$. 
Such an object may also, in principle,  carry some number of Lorentz indices, 
but here we are interested in its SU(2)-structure only.
The  gauge transformation law of $F^{i_1 \dots i_n}$ is given by eq. 
(\ref{tensor-K}). In particular, the local SU(2) transformation, 
which is described by  parameters $K^{ij} =K^{ji}$,
acts on $F^{i_1 \dots i_n} $ as follows:
\bea
\d_{\rm SU(2)} F^{i_1\cdots i_n} 
\equiv K^{kl} J_{kl}  \,F^{i_1\cdots i_n} 
=
\sum_{l=1}^{n} K^{i_l}{}_j \,F^{j i_1\cdots  \widehat{i_l} \cdots i_n} 
=n  \,F^{j (i_1 \cdots i_{n-1}} K^{i_n)}{}_j 
~,
\label{F-K}
\eea
where the notation $\widehat{i_k} $ means that the corresponding index is missing.

It is useful to develop an alternative description for the above superfield 
as a holomorphic tensor field over ${\mathbb C}P^1$.
With the aid of complex variables
$u^+_i \in {\mathbb C}^2 \setminus \{0\}$, 
following \cite{Waerden},
let us associate 
with $F^{i_1 \dots i_n}(z)$  
a  homogeneous polynomial of $u^+$ of degree $n$
defined as
\bea
F^{(n)}(z,u^+)= u^+_{i_1}\cdots u^+_{i_{n}}\,F^{i_1\cdots i_{n}}(z)~, 
\qquad F^{(n)}(z,c\,u^+)=c^n \,F^{(n)}(z,u^+)~.
\label{F-n}
\eea
It is convenient to interpret the variables $u^+_i$ to be homogeneous coordinates
for ${\mathbb C}P^1$. The latter space 
emerges by factorizing 
${\mathbb C}^2 \setminus  \{0\}$ with respect to the equivalence 
relation $u^+_i \sim c\,u^+_i$, with $c\in \mathbb{C}^*$.
Then, $F^{(n)}$ is known to define a holomorphic tensor field
of rank $(n/2, 0)$ on  ${\mathbb C}P^1$.
Eq. (\ref{F-K}) can now be interpreted as a transformation  acting 
in the space of holomorphic tensor fields
of rank $(n/2, 0)$ on  ${\mathbb C}P^1$.
It is  defined as 
\bea
\d_{\rm SU(2)}  F^{(n)}(z,u^+)
&:=&
 u^+_{i_1}\cdots u^+_{i_{n}}\,\d_{\rm SU(2)}F^{i_1\cdots i_{n}} (z)~.
\label{F-K-u}
\eea
It turns out that this transformation law 
can be rewritten as follows:
\bea
\d_{\rm SU(2)}F^{(n)}
\equiv K^{kl} J_{kl}  \,F^{(n)}
&=&  -\frac{1}{(u^+u^-)} \Big(K^{++} D^{--} 
-n \, K^{+-}\Big) F^{(n)} ~, 
\label{F-K2} \\
 K^{\pm \pm } &=&K^{ij}\, u^{\pm}_i u^{\pm}_j ~, 
\non
\eea 
with the first-order operator $D^{--}$  defined in (\ref{5}).
The right-hand side in (\ref{F-K2})
involves an auxiliary complex two-vector
$u^-_i$ which 
is chosen to be linearly independent of $u^+_i$, 
that is $(u^+u^-) := u^{+i}u^-_i \neq 0$, but is otherwise completely arbitrary.
By construction, both $F^{(n)}$ and $\d_{\rm SU(2)} F^{(n)}$ are independent of $u^-$.
It should be pointed out that eq. (\ref{F-K2}) defines the action of 
the covariant derivatives $\cD_A$, eq. (\ref{CovDev}), on $F^{(n)}$
(for any super-vector field $\x^A(z)$, the operator $\x^A \,\cD_A$
acts on the space of superfields $F^{(n)}$).

If  there are
two homogeneous polynomials  $F^{(n)}(u^+)$ and $F^{(m)}(u^+)$, 
their product $F^{(n+m)}(u^+):=F^{(n)}(u^+)\,F^{(m)}(u^+)$
is a homogeneous polynomials of order $(n+m)$. 
In superspace, new covariant operations can be defined.
Indeed, one can  allow the polynomials $F^{(n)}(u^+)$ to be tensor 
superfields, i.e.  be $z$-dependent and carry Lorentz indices. 
Then, the spinor covariant derivatives can be used to define covariant  maps 
of $F^{(n)}$'s to  $F^{(n+1)}$'s by the rule:
\begin{subequations}
\bea
\cD^+_{ \a} F^{(n)} (z,u^+)&:= &
u^+_j u^+_{i_1}\cdots u^+_{i_{n}}\,\cD^j_\a F^{i_1 \cdots i_{n}} (z)
=u^+_j u^+_{i_1}\cdots u^+_{i_{n}}\,\cD^{(j}_\a F^{i_1 \cdots i_{n})} (z)~,
\label{def-D}\\
{\bar \cD}^+_{ \ad} F^{(n)} (z,u^+)&:=& 
u^+_j u^+_{i_1}\cdots u^+_{i_{n}}\,{\bar \cD}^j_\ad F^{i_1 \cdots i_{n}} (z)
=u^+_j u^+_{i_1}\cdots u^+_{i_{n}}\,{\bar \cD}^{(j}_\ad F^{i_1 \cdots i_{n})} (z)~.
\label{def-bar-D}
\eea
\end{subequations}
The superfield $\cD^+_{ \a} F^{(n)}$ and  ${\bar \cD}^+_{ \ad} F^{(n)}$
obtained are  
of the type 
$F^{(n+1)}$. 
Therefore, the operators $\cD^+_{ \a}$ and   
${\bar \cD}^+_{\dot  \a}$ are covariant derivatives 
that send $F^{(n)}$'s to  $F^{(n)}$'s.
With the definitions $\cD^+_{ \a}:=u^+_j\,\cD^j_{ \a} $ and 
$\cD_{\a}^j=E_{\a}^j+  \hf \,\O_{\a}^j{}^{bc}\,M_{bc} +
\Phi_{\a}^j{}^{kl}\,J_{kl}$, the right-hand side in  (\ref{def-D}) is
actually 
a direct consequence of (\ref{F-K-u}).

When acting on $ F^{(n)}$,  
the operators $\cD^+_{ \a}$ and   
${\bar \cD}^+_{\dot  \a}$ can be seen to 
obey the anticommutation relation  (\ref{analyt}).
For instance, it follows from the definition (\ref{def-D})
\bea
\{\cD^+_\a , \cD^+_\b \} F^{(n)} 
= u^+_j u^+_k u^+_{i_1}\cdots u^+_{i_{n}}\,
\{ \cD^{(j}_\a , \cD^k_\b \} F^{i_1 \cdots i_{n})} ~,
\eea
and it only remains to apply (\ref{acr1}). 
Recalling the explicit action of the SU(2) generators on isospinors, 
eq. (\ref{generators}), 
for the operator $J^{++}:=u^+_j u^+_k J^{jk}$
appearing in (\ref{analyt}) one obtains
\be
J^{++}F^{(n)} = u^+_j u^+_k u^+_{i_1}\cdots u^+_{i_{n}}\,
J^{ (jk} F^{i_1 \cdots i_{n})} =0~.
\ee

In accordance with the definition of $\d_{\rm SU(2)}F^{(n)}(z,u^+)$, 
eq. (\ref{F-K-u}), the auxiliary coordinates $u^+_i$ 
are inert under the local SU(2) transformations, $\d_{\rm SU(2)} u^+_i=0$. 
This is similar to the point  of view  adopted for 
the superspace coordinates $z^M$.
These variables are chosen to be inert under the supergravity 
gauge transformations (\ref{tau}) and (\ref{tensor-K}). 
The latter transform only the functional form of the 
dynamical superfields.
Since $u^+_i$ are inert under the local SU(2) transformations, 
these variables are covariantly constant, $\cD_A u^+_i=0$.
The latter property is 
implied by 
eqs.  (\ref{def-D}) and (\ref{def-bar-D}) in conjunction with the Leibniz rule.

The example of $F^{(n)}$'s considered
can naturally be extended to define more general 
 superfields.  
Let us consider a superfield  $U^{(n)} (z,u^+) $
(with its Lorentz indices suppressed)   
that
lives on  $\cM^{4|8}$, 
is holomorphic with respect to 
the isotwistor variables $u^{+}_i $ on an open domain of 
${\mathbb C}^2 \setminus  \{0\}$, 
and is characterized by the following conditions:\\
(i) it is  a homogeneous function of $u^+$ 
of degree $n$, that is,  
\be
U^{(n)}(z,c\,u^+)\,=\,c^{n}\,U^{(n)}(z,u^+)~, \qquad c\in \mathbb{C}^*~;
\label{U-homog}
\ee
(ii)  supergravity gauge transformations (\ref{tau}) act on $U^{(n)}$ 
as follows:
\bea
\d_K U^{(n)} 
&=& \Big( K^{{C}} \cD_{{C}} + \hf K^{cd}M_{cd} +K^{ij} J_{ij}  \Big)U^{(n)} ~,  
\non \\ 
K^{ij} J_{ij}  \,U^{(n)}&=& -\frac{1}{(u^+u^-)} \Big(K^{++} D^{--} 
-n \, K^{+-}\Big) U^{(n)} ~. 
\label{U-K}
\eea 
The latter  relation also defines  the  action of the 
covariant derivative
$\cD_{{A}}$, eq. (\ref{CovDev}), 
 on $U^{(n)}(z,u^+)$.
By construction, $U^{(n)}$ is independent of $u^-$, 
i.e. $\pa  U^{(n)} / \pa u^{-i} =0$,
hence $D^{++}U^{(n)}=0$.
One can  check
that $\d_K U^{(n)} $ 
is also independent of 
$u^-$, $\pa (\d_K U^{(n)})/\pa u^{-i} =0$, 
as a consequence of (\ref{U-homog}).
Defining
\be
J^{++} = u^+_i u^+_j J^{ij}~, \qquad 
J^{+-} = u^+_i u^-_j J^{ij}~,
\ee
eq. (\ref{U-K}) implies
\bea
J^{++} \,U^{(n)}=0~, \qquad 
 J^{+-} \,U^{(n)}=-\frac{n}{2}\, (u^+u^-) \,U^{(n)}~.
\eea
We will call $U^{(n)} (z,u^+) $ an isotwistor superfield 
of weight $n$.

Now, consider the  covariant derivatives $\cD^+_{ \a}:=u^+_i\,\cD^i_{ \a} $ and   
${\bar \cD}^+_{\dot  \a}:=u^+_i\,{\bar \cD}^i_{\dot \a} $. 
It is evident  that  $\cD^+_{ \a}U^{(n)}$ 
and ${\bar \cD}^+_{\dot  \a}U^{(n)}$ are isotwistor superfields 
of weight $(n+1)$.
When acting on isotwistor superfields, 
the operators $\cD^+_{ \a}$ and   
${\bar \cD}^+_{\dot  \a}$
obey the anticommutation relation  (\ref{analyt}).

\end{appendix}

\small{

}

\end{document}